\documentstyle[aps,prd,preprint,epsfig]{revtex}

\begin{document}

\title{A Model of Charmonium Absorption by Light Mesons}

\author{Sergei G. Matinyan and Berndt M\"uller\\
Department of Physics, Duke University, Durham, NC 27708-0305}

\date{\today}

\maketitle

\begin{abstract}
We calculate the cross sections for dissociation of $J/\psi$ by pions
and $\rho$-mesons within the framework of a meson exchange model. We
find that these cross sections are small at center-of-mass energies
less than 1 GeV above threshold, and that dissociation rates are
less than 0.01 $c$/fm in a thermal meson gas at temperatures where
such a description makes sense.
\end{abstract}


A pronounced suppression of the production of strongly bound heavy quarkonium
states ($J/\psi$, $\Upsilon$) is considered as a signature for the formation
of a new, deconfined phase of strongly interacting matter in collisions
of heavy nuclei at high energy, the quark-gluon plasma \cite{MS85}.
Experimental investigations of charmonium production in nuclear reactions
have been carried out for over a decade at the CERN-SPS (p+A, O+U, S+U,
Pb+Pb) and the Fermilab Tevatron (p+A) \cite{E772,NA38,NA38p,NA50}. 
These studies have shown that the
$J/\psi$ and $\psi'$ production on nuclear targets is, indeed suppressed
relative to expectations from nucleon-nucleon reactions. 

For p+A collisions the observed suppression of $J/\psi$ and $\psi'$ 
can be explained as absorption of a common precursor, probably 
a non-resonant, color-octet $(c\bar{c})$ state, on nucleons \cite{KS96}.
This systematics extends to O+U and S+U reactions for the $J/\psi$, but
not for the $\psi'$, where additional suppression is observed \cite{NA38p}.
This effect can be quantitatively described as absorption of the $\psi'$
on comoving secondary hadrons (``comovers''), mostly light mesons 
\cite{KLNS97,GV96,Wong96,CKKG96}.
Recently, measurements of charmonium production in the Pb+Pb system
\cite{NA50} have revealed the presence of an additional ``anomalous''
suppression mechanism also for the $J/\psi$. It is presently under debate 
whether this mechanism can be absorption of the $J/\psi$ or the down-feeding
$\chi_c$ on hadronic comovers, or whether a more drastic explanation,
such as quark deconfinement, is required \cite{AC97,Vogt97,KNS97}.

In spite of the obvious need to understand the physics of charmonium
absorption on comoving hadrons, only very few quantitative predictions
for $J/\psi$ absorption cross sections on light mesons have been
published \cite{KS94,MBQ}., and these predictions differ by several orders of
magnitude. We hope that the results presented below will help clarifying
this situation. Our calculation is based on an effective hadronic 
Lagrangian describing the interactions among $\pi$, $\rho$, $J/\psi$,
$D$, and $D^*$ and, for the sake of simplicity, on the assumption of 
thermal equilibrium among the light hadrons.

The most abundant mesons in a hot hadronic gas are $\pi$, $K$, and
$\rho$.  In addition, nucleons may be abundant at high baryon
chemical potential.  In the following, we will not consider kaon
induced reactions, but the framework presented here can be easily 
extended to include this absorption channel.  At low energy, $\pi$ 
and $\rho$ mesons, as well as nucleons, can induce dissociation 
processes when encountering a $J/\psi$ particle
(we list only the reactions with the lowest energy thresholds):
\begin{eqnarray}
\pi + J/\psi &\to & D+ \bar D^*, \quad \bar D + D^* ; \label{e21} \\
\rho + J/\psi &\to & D+\bar D; \label{e23} \\
N + J/\psi &\to & \Lambda_c + \bar D. \label{e22}
\end{eqnarray}
The kinematic thresholds for these reactions are (\ref{e21}) 640 MeV,
(\ref{e23}) $-135$ MeV, and (\ref{e22}) 115 MeV, respectively.
The reaction $\rho+J/\psi\to D+\bar D$ has the lowest total invariant
mass threshold and is, in fact, exothermic.  At thermal equilibrium it
is only suppressed due to the relatively high mass of the
$\rho$-meson which causes $\rho$-mesons to be less abundant than pions 
in a thermal hadron gas. The reaction (\ref{e22}) is mostly of interest
in the fragmentation region, where the nucleon density is high. We
will not consider it further here, because it is expected to be
of less importance for $J/\psi$ suppression at central rapidity

There are two possible approaches to the $J/\psi$ dissociation problem:  
one at the quark level where the large mass of the $c$-quark is used to
separate perturbative from nonperturbative aspects of the problem, and
another one that makes use of effective hadronic interactions.  
The former approach is based on the 
pioneering work of Peskin \cite{Pes79} and Bhanot and Peskin \cite{BP79}, 
who realized that interactions between heavy quark bound states and
light hadrons can be described perturbatively, if the heavy quark mass
$m_Q$ is sufficiently large.  
The small size of the $(Q\bar Q)$ state allows for a
systematic multipole expansion of its interaction with external glue
fields, where the color-dipole interaction dominates at long range.

In the Bhanot-Peskin framework, light hadrons interact with the $J/\psi$
only via their glue content.  Kharzeev and Satz \cite{KS94} applied this
formalism to inelastic $\pi-J/\psi$ scattering and showed
that the absorptive cross section is proportional
to the component of the gluon structure function $G_h(x)$ of the
interacting hadron which is sufficiently energetic to dissociate the
$J/\psi$.  Because of this requirement of rather hard gluons, 
only highly energetic hadrons are capable of exciting a $J/\psi$ above 
the dissociation threshold.  The pion momentum must reach 
5 GeV/$c$ before attaining an absorption cross section in excess of 1 mb.  
The thermally averaged cross section, in this framework, remains less than 
0.1 mb for a pion gas within any realistic temperature range.

One limitation of the Bhanot-Peskin approach lies in the fact that 
the $J/\psi$ is not truly a Coulombic bound state but probes also the 
confining part of the $c\bar c$-potential.  Although these nonperturbative 
contributions to $J/\psi$ absorption by light mesons can be estimated
\cite{KMS95}, the reliability of such estimates is difficult to assess.

From a microscopic point of view, the reactions (\ref{e21},\ref{e23}) 
can be viewed as quark exchanges, where the $J/\psi$ transmits a charm
quark to the light meson and picks up a light $(u,\;d,\;{\rm or}\;s)$
quark.  Since similar reactions among light hadrons typically have
large cross sections at moderate energies, one may suspect that these 
reactions also proceed with significant strength above their respective 
kinematic thresholds. 
It is therefore of interest to calculate the dissociation cross
sections in the framework of an essentially nonperturbative approach
based on a hadronic model that incorporates quark confinement.  
The cross section for the reaction (\ref{e21}) was calculated 
in the nonrelativistic quark model by Martins, Blaschke, and Quack 
\cite{MBQ} in the first Born approximation.
Including $D^*\bar D,\; D\bar D^*$, and $D^*\bar D^*$ final states,
the total cross section was found to peak around 1 GeV energy above threshold
(in the center-of-mass system) at a value of about 7 mb.
However, the large magnitude of their result
is due to the action of the long-ranged, confining interaction between
the quarks, which is modeled as a ``color-blind'' attractive interaction
between the quarks with a Gaussian momentum dependence. Because
this interaction is taken as attractive independent of the color
quantum numbers of the affected quark pair, it does not cancel in
the interaction of a light quark with a point-like ($c\bar{c}$) pair
in a color-singlet state, in contrast to the one-gluon exchange
interaction.  

Since the results obtained within the constituent quark model differ by
orders of magnitude from those obtained in the Bhanot-Peskin approach,
and because they depend critically on the particular implementation
of the confining quark-quark interaction, it seems prudent to calculate 
the charm-exchange cross sections in a second, entirely different 
framework of hadronic reactions.  In the effective meson theory,
the exchange of a $(c\bar q)$ or $(\bar cq)$ pair, where $q,\bar q$
stands for any light quark, can be described as the exchange of a $D$
or $D^*$ meson between the $J/\psi$ and the incident light meson.  

At the hadronic level, the charm exchange reaction between the
$J/\psi$ and a light hadron can proceed either by exchange of a $D$-
or a $D^*$-meson.  In fact, Regge theory dictates that the charm exchange 
reaction is dominated by the exchange of the $D^*$-trajectory in the 
high energy limit.  However, here we are not interested in charm
exchange at high energies but near the kinematical threshold, because
the relative motion of the hadrons is limited to thermal momenta.  
We shall only consider the $D$-exchange reactions.
(The $D^*$-exchange reaction cross sections exhibit an unphysical
rise with energy due to the exchange of longitudinally polarized 
$D^*$-mesons, which would need to be eliminated with the help of
form factors.)

In order to calculate the various Feynman diagrams for the 
reactions (\ref{e21}--\ref{e23}) we need to construct the effective 
three-meson vertices.  We can do so by invoking a strongly broken local 
U(4) flavor symmetry with the vector mesons playing the role of quasi-gauge 
bosons.  Denoting the 16-plet of pseudoscalar mesons $(\pi,\eta,\eta',
K,D,D_s,\eta_c)$ by $\Phi=\phi_iT_i$, where the $T_i$ are the $U(4)$ 
generators, and the vector meson 16-plet $(\omega,\rho,\phi,K^*,D^*,D_s^*,
\psi)$ by ${\cal V}^{\mu} = V_i^{\mu}T_i$, the free meson Lagrangian reads
\begin{equation}
{\cal L}_0 = {\rm tr} (\partial^{\mu}\Phi^{\dagger}\;
\partial_{\mu}\Phi) - {\rm tr} (\partial^{\mu}{\cal
V}^{\dagger,\nu})(\partial_{\mu}{\cal V}_{\nu}-\partial_{\nu}{\cal
V}_{\mu}) \quad - {\rm tr} (\Phi^{\dagger}M_{\rm P}\Phi) + {1\over 2}
{\rm tr}({\cal V}^{\mu_{\dagger}}M_{\rm V}{\cal V}_{\mu}). \label{e13}
\end{equation}
Here $M_{\rm P}$ and $M_{\rm V}$ denote the mass matrices for the 
pseudoscalar and vector mesons, respectively.  Because of the heavy mass 
of the charm quark, $M_{\rm P}$ and $M_{\rm V}$ break the U(4) symmetry 
strongly down to U(3), the mass of the strange quark introduces a weaker 
breaking to U(2), and the axial anomaly further breaks the U(2) symmetry 
to SU(2) in the case of the pseudoscalar mesons.  All these symmetry
breakings are embodied in the physical mass matrices.  It is
convenient, in the following, to work with the mass eigenstates.

The meson couplings are obtained by replacing the derivatives 
$\partial_{\mu}$ by the ``gauge covariant'' derivatives
$D_{\mu} = \partial_{\mu} - ig {\cal V}_{\mu}$.
In first order in the coupling constant $g$ this procedure leads to
the following interactions
\begin{equation}
{\cal L}_{\rm int} = ig\; {\rm tr} \left( \Phi^{\dagger} {\cal
V}^{\mu \dagger}\partial_{\mu}\Phi - \partial^{\mu}\Phi^{\dagger}{\cal
V}_{\mu}\Phi\right) 
+ ig\; {\rm tr} \left( \partial^{\mu}{\cal V}^{\dagger\nu}
[{\cal V}_{\mu},{\cal V}_{\nu}] - [{\cal V}^{\dagger \mu},{\cal
V}^{\dagger \nu}]\partial_{\mu}{\cal V}_{\nu}\right). \label{e15}
\end{equation}
If the U(4) flavor symmetry were exact, we would expect all
couplings given by the same constant $g$.  In view of the significant
breaking of the flavor symmetry we anticipate that the effective
coupling constants for different 3-meson vertices will have different
values.  However, we will see below that the coupling have remarkably
similar values.

In order to describe $\pi$- and $\rho$- induced $J/\psi$ dissociation,
we need the $\psi DD$, $\pi DD^*$, and $\rho DD$ vertices.  
From (\ref{e15}) we derive the following interactions:
\begin{eqnarray}
{\cal L}_{\pi DD^*} &= &{i\over 2}g_{\pi DD^*} \left( \bar D \tau_i
D^{*\mu}\partial_{\mu}\pi_i - \partial^{\mu} D \tau_i D_{\mu}^*
\pi_i - {\rm h.c.} \right) \nonumber \\
{\cal L}_{\rho DD} &= &{i\over 2}g_{\rho DD} \rho_i^{\mu} \left( \bar D
\tau_i \partial_{\mu} D - \partial_{\mu}\bar D \tau_i D \right) \nonumber \\
{\cal L}_{\psi DD} &= &ig_{\psi DD} \psi^{\mu}\left( \bar
D\partial_{\mu}D - (\partial_{\mu}\bar D) D\right).
\label{e16}
\end{eqnarray}
Here the $\tau_i$ denote the generators of SU(2)-isospin symmetry
in the fundamental representation (Pauli matrices).
The coupling constants $g_{\psi DD}$, and $g_{\rho DD}$
can be derived from the $D$-meson electric form factor in the 
standard framework of the vector meson dominance model \cite{VMD}.
If $\gamma_{\rm V}$ denotes the photon-vector meson ${\cal V}$ mixing
amplitude, the usual analysis yields the relations
\begin{equation}
\gamma_{\rho}g_{\rho DD} = em_{\rho}^2, \qquad \gamma_{\psi} g_{\psi DD}
= {2\over 3}em_{\psi}^2 \label{e17}
\end{equation}
The photon mixing amplitudes $\gamma_{\rm V}$ can be determined from the 
leptonic vector meson decay widths:
$\Gamma_{\rm Vee} = \alpha \gamma_{\rm V}^2 / (3 m_{\rm V}^3)$.
Inserting the experimental numbers, one finds
\begin{equation}
g_{\rho DD} \approx 5.6, \qquad g_{\psi DD} \approx 7.7. \label{e20}
\end{equation}
As is well known, different ways of deriving the value of the
$\rho$-meson coupling yield values differing by about 20\%.  
The coupling constants (\ref{e20}) must therefore be
considered to be given with an error of this order of magnitude.

The $\pi$-meson coupling between $D$ and $D^*$ can be
obtained from the decay width of the $D^*$-meson:  $D^*\to D\pi$.
Unfortunately, only an upper bound for this decay rate is known at
present \cite{Bar92}, corresponding to $g_{\pi DD^*}<15$.  
A theoretical estimate for this decay rate, based on QCD sum rules 
\cite{Bel95}, gives
\begin{equation}
g_{\pi DD^*} \approx 8.8.
\end{equation}
For lack of an experimentally determined value of $g_{\pi DD^*}$ we 
adopt this value here.

\begin{figure}
\centerline{\mbox{\epsfig{file=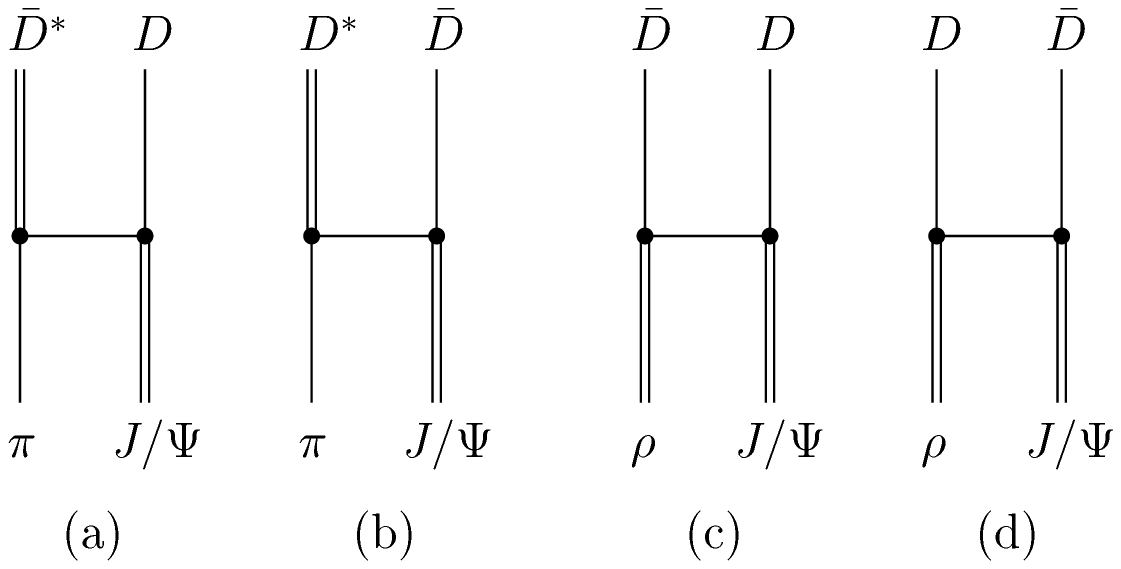,width=.9\linewidth}}}
\vspace{2truein}
\caption{Lowest order Feynman diagrams contributing to the charm
exchange reactions \protect (\ref{e21}) and \protect (\ref{e23}).
Single lines represent pseudoscalar mesons; double lines denote
vector mesons.}
\label{fig9}
\end{figure}

Analytical expressions for the amplitudes and cross sections for the
processes corresponding to the Feynman diagrams in Figure \ref{fig9} are
easily derived.  We begin with the two diagrams (\ref{fig9}a,b) for 
absorption by pions.  Averaging over initial and summing over final 
isospin and spin states, one obtains
\begin{equation}
\overline{\vert M_a\vert^2} = {1\over 6} g_{\psi DD}^2 g_{\pi DD^*}^2
{1\over t^{'2}} \left( 4m_{\pi}^2 - {(t-m_{D^*}^2-m_{\pi}^2)^2 \over
m_{D^*}^2}\right) 
\left( 4m_D^2 - {(t-m_D^2-m_{\psi}^2)^2 \over m_{\psi}^2}\right) 
= \overline{\vert M_b\vert^2} . \label{B2}
\end{equation}
Here $s$, $t$, and $u$ are the Mandelstam variables and we have
introduced the notation $t'=t-m_D^2$, $u'=u-m_D^2$.
The contributions of the diagrams (\ref{fig9}c,d) need to be added coherently,
because they lead to identical final states.  However, their results are
related by crossing symmetry $(t\leftrightarrow u)$.  After some
algebra one finds:
\begin{eqnarray}
\overline{\vert M_c+M_d\vert^2} &=& \frac{1}{18} g_{\psi DD}^2 
g_{\rho DD}^2 \left[ \frac{1}{t^{'2}} \left( 4m_D^2 -
\frac{(m_{\rho}^2-t')^2}{m_{\rho}^2}\right) \right.
\left( 4m_D^2 - \frac{(m_{\psi}^2-t')^2}{m_{\psi}^2}
\right)  \nonumber \\
&&+ \frac{1}{u't'} \left( 2s-4m_D^2 -
\frac {(m_{\rho}^2-t')(m_{\rho}^2-u')}{m_{\rho}^2} \right) 
\left(2s - 4m_D^2 - \frac{(m_{\psi}^2 - t')(m_{\psi}^2-u')}{m_{\psi}^2}
\right) \nonumber \\
&&+ \left. (t' \leftrightarrow u') 
\phantom{\frac{m_{\psi}^2}{m_{\psi}^2}} \right]. \label{B4}
\end{eqnarray}  

\begin{figure}
\centerline{\mbox{\epsfig{file=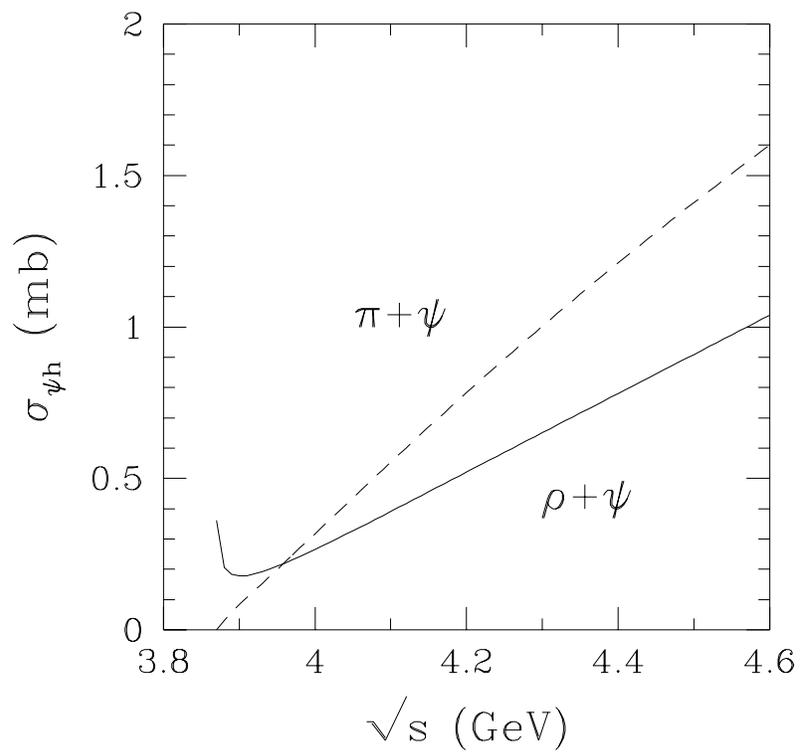,width=.7\linewidth}}}
\vspace{2truein}
\caption{Cross sections for the charm exchange reactions described by
the diagrams of Figure \protect \ref{fig9}, as functions of c.m.
energy. Dashed line: pions, solid line: $\rho$-mesons.}
\label{fig10}
\end{figure}

The charm exchange cross sections are plotted in Figure \ref{fig10} 
as functions of the center-of-mass energy $\sqrt{s}$.  
The $\pi +J/\psi$ cross section (dashed line) starts at
zero, because the reaction is endothermic, whereas the $\rho+ J/\psi$
cross section (solid line) is finite at the threshold because of its
exothermic nature.  Note that the cross sections for these two $J/\psi$ 
absorption reactions are of similar magnitude over the energy
range relevant to a thermal meson environment.

\begin{figure}
\centerline{\mbox{\epsfig{file=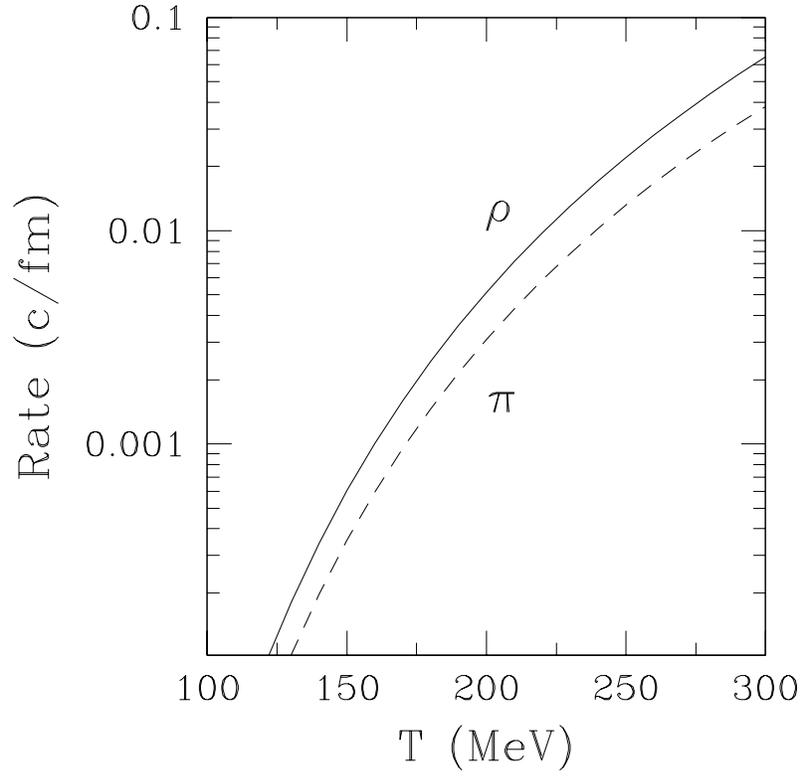,width=.7\linewidth}}}
\vspace{2truein}
\caption{Thermal $J/\psi$ absorption rates as function of temperature.
The pion and $\rho$-meson rates are shown separately.}
\label{fig12}
\end{figure}

Although pions are far more abundant than $\rho$-mesons in the temperature 
range where a hadronic gas is likely to exist, most of these pions do
not have sufficient energy to dissociate a $J/\psi$.  
If one counts only those pions having energy
above the kinematical threshold for $J/\psi$-dissociation, $\pi$- and
$\rho$-mesons are about equally rare, because their effective density 
not exceeding 0.1 fm$^{-3}$ even at $T = 200$ MeV, where the hadron 
picture of a thermal environment probably already breaks down. As a
consequence, we expect dissociation rates of similar size for $\pi$
and $\rho$ mesons at thermal equilibrium.  Figure \ref{fig12} shows 
these $J/\psi$ absorption rates as a function of temperature.  
Even at the (unrealistically) high temperature $T = 300$ MeV, the 
thermal dissociation rates are still so small that a $J/\psi$
continues to have a lifetime around 10 fm/$c$.  For $T \le$ 200 MeV,
dissociation by hadrons is completely negligible on the time scale 
of the lifetime of a hot hadronic gas state in nuclear collisions.

The inclusion of a form factor in the meson vertices of the Feynman 
diagrams (\ref{fig9}) would further reduce the dissociation rates.
E.g. a Gaussian form factor $\exp(-Q^2/Q_0^2)$ with $Q_0 = 1.5$ GeV 
would reduce the absorption rates by slightly more than one order of 
magnitude.  This implies that our result represents an upper limit
for the hadronic dissociation rate.  As we noted above, our analysis 
remains incomplete because of the neglect of $D^*$-exchange reactions.  
It would be most interesting to include these reactions in the framework 
of a complete effective Lagrangian describing the interactions of the
pseudoscalar and vector mesons with quark content $(q\bar q)$, $(Q\bar q)$,
$(q\bar Q)$, and $(Q\bar Q)$, where $q$ stands for any light quark flavor 
and $Q$ denotes a heavy quark.  Such a Lagrangian embodying chiral 
symmetry for the light quarks has recently been given by Chan \cite{Chan}. 
Loop corrections to the tree diagrams would permit the study of form
factor effects in this approach, as well.

In conclusion, we have calculated the dissociation cross sections and
thermal absorption rates of $J/\psi$ mesons on pions and $\rho$-mesons
in the framework of an effective meson exchange model. We find that
the dissociation rates in a thermal meson gas, at temperatures where such
a gas is expected to exist, are very small corresponding to a survival
time of the $J/\psi$ in excess of 100 fm/$c$. Our results are in line
with those recently obtained for other absorption channels, such as
$\pi + J/\psi \to \pi + \psi'$ \cite{CS97} and $\rho + J/\psi \to
\pi + \eta_c$ \cite{ST98}. Taken together, these results indicate that
a thermal hadronic environment in the confined phase of QCD cannot 
be the cause of significant absorption of $J/\psi$ mesons.

{\it Acknowledgments:}
This work was supported in part by a grant from the U.S.
Department of Energy (DE-FG02-96ER40945).

\newpage

\newpage

\newpage

\newpage

\end{document}